\pdfoutput=1 
\documentclass{article}
\usepackage{siunitx}

\usepackage[utf8]{inputenc}
\usepackage[margin=0.75in]{geometry} %
\usepackage{authblk} %
\usepackage{lineno}

\usepackage{newunicodechar}

\usepackage[backend=biber,style=nature,sorting=none]{biblatex} %
\let\cite=\supercite

\usepackage{algorithm,amsmath}

\usepackage{hyperref} 
\usepackage{cleveref}
\usepackage{xcolor}

\hypersetup{
  colorlinks   = true, %
  urlcolor     = blue, %
  linkcolor    = blue, %
  citecolor   = blue %
}

\usepackage[noend]{algpseudocode}
\usepackage{graphicx}
\usepackage[labelformat=simple]{subcaption} 
\usepackage[hypcap=true]{caption}
\usepackage[T1]{fontenc}
\usepackage{lmodern}
\usepackage[labelfont=bf]{caption}

\title{Effective approaches to disaster evacuation during a COVID-like pandemic}

\author[1,*]{Yi-Lin Tsai}
\author[2]{Dymasius Y. Sitepu}
\author[3]{Karyn E. Chappell}
\author[4]{Rishi P. Mediratta}
\author[4,5]{C. Jason Wang}
\author[1,6,7,8]{Peter K. Kitanidis}
\author[6,9,10,11]{Christopher B. Field}

\affil[1]{Department of Civil and Environmental Engineering, Stanford University, Stanford, CA, USA}
\affil[2]{Department of Engineering Science, National University of Singapore, Singapore}
\affil[3]{Department of Engineering, Imperial College London, London, UK}
\affil[4]{Department of Pediatrics, Stanford University School of Medicine, Stanford, CA, USA}
\affil[5]{Department of Health Policy, Stanford University School of Medicine, Stanford, CA, USA}
\affil[6]{Woods Institute for the Environment, Stanford University, Stanford, CA, USA}
\affil[7]{Bio-X, Stanford University, Stanford, CA, USA}
\affil[8]{Institute for Computational and Mathematical Engineering, Stanford University, Stanford, CA, USA}
\affil[9]{Department of Biology, Stanford University, Stanford, CA, USA}
\affil[10]{Department of Earth System Science, Stanford University, Stanford, CA, USA}
\affil[11]{Interdisciplinary Environmental Studies Program, Stanford University, Stanford, CA, USA}

\affil[*]{yilin2@stanford.edu}

\begin{document}

\date{\today}
\vspace{-0.5in}

\maketitle

\begin{abstract}

Since COVID-19 vaccines became available, no studies have quantified how different disaster evacuation strategies can mitigate pandemic risks in shelters. Therefore, we applied an age-structured epidemiological model, known as the Susceptible-Exposed-Infectious-Recovered (SEIR) model, to investigate to what extent different vaccine uptake levels and the Diversion protocol implemented in Taiwan decrease infections and delay pandemic peak occurrences. Taiwan’s Diversion protocol involves diverting those in self-quarantine due to exposure, thus preventing them from mingling with the general public at a congregate shelter. The Diversion protocol, combined with sufficient vaccine uptake, can decrease the maximum number of infections and delay outbreaks relative to scenarios without such strategies. When the diversion of all exposed people is not possible, or vaccine uptake is insufficient, the Diversion protocol is still valuable. Furthermore, a group of evacuees that consists primarily of a young adult population tends to experience pandemic peak occurrences sooner and have up to 180\% more infections than does a majority elderly group when the Diversion protocol is implemented. However, when the Diversion protocol is not enforced, the majority elderly group suffers from up to 20\% more severe cases than the majority young adult group.

\end{abstract}
 

\section*{Introduction}
The best tools for minimizing the number of infections are social distancing, wearing masks, regular testing, isolation, quarantine, contact tracing, and vaccination. However, these tools are less effective when a natural disaster or a humanitarian crisis occurs, especially when many people in a community without population-wide COVID-19 contact tracing and testing must evacuate in a chaotic, time-constrained setting. Evacuees may not be able to consistently wear masks, and the asymptomatic infected evacuees may cause COVID-19 outbreaks both during the evacuation and while evacuees stay at shelters. Thus, potential outbreak risks are embedded in cascading disasters when a pandemic overlaps with a natural hazard, such as a debris flow, or a humanitarian crisis, such as a war.

Public policy frameworks and recommendations for planning disaster evacuations (\textit{e.g.}, floods, fires, earthquakes, and hurricanes) during the COVID-19 pandemic have proliferated \cite{simonovic2021floods, ashraf2021lessons, rahman2021covid, james2020cascading, shultz2020mitigating, ishiwatari2020managing, montz2020risk, markotic2020earthquake, sakamoto2020implementation}. However, to the best of our knowledge, this is the first study to investigate the interactive effects of vaccination rate and evacuation strategy on disease transmission and dynamics during a compound disaster.

This study aims to save lives in future infectious-disease outbreak responses and effectively help governments decide on the optimal response to reduce virus transmission during disaster evacuation by analyzing real-world data from Taiwan of two vaccine uptake levels. Currently, more infectious SARS-CoV-2 variants, like alpha, beta, gamma, delta, omicron, and BA.2, have emerged \cite{rahimi2022omicron, karim2021omicron}. In addition to the public health risks caused by a prolonged pandemic, more severe hurricane and wildfire seasons have occurred in recent years, and earthquakes can occur anytime. Recent disaster events or humanitarian crises around the world have complicated evacuations. These complicated evacuations are evident in the case of the deadly floods and debris flows in Japan, Germany, Belgium, and China, Super Tropical Storm Amphan in South Asia, and a dam collapse in Michigan, USA. These catastrophes worsen COVID-19 outbreaks and subsequent socio-political consequences.

For an example of compound crises, we can look to Super Tropical Storm Amphan, which hit India and Bangladesh in May 2020. The governments of these countries ordered millions of people to evacuate to congregate shelters, where they could not maintain social distancing and did not wear masks \cite{pramanik2021population, islam2021revisiting}. Instead of complying with these mandatory evacuation orders, however, many residents in both countries chose to stay at home because they did not want to expose themselves to the risks of COVID-19 infections \cite{tozier2021managing}. Thus, some people's hesitation to evacuate caused inefficiencies in disaster response and recovery. This kind of indecision over evacuation and shelter in place might decrease COVID-19 cases, but at the cost of increased fatalities and property loss. There is an urgent need to evaluate and eventually implement early preventative measures and coherent protocols when COVID-19 coincides with other disasters \cite{ishiwatari2020managing}.

Although some studies have provided quantitative analyses, these were done before COVID-19 vaccines were available. Several of these studies focused on the effects of different evacuation strategies on the consequences of natural hazards before and after the COVID-19 pandemic. For example, Dutch researchers analyzed the consequences of various evacuation timelines and strategies for hypothetical large-scale flooding in the Netherlands \cite{kolen2012time}. Other studies used epidemiological models to simulate hypothetical \cite{tripathy2021flood, pei2020compound} and historical \cite{quigley2020multi} evacuation scenarios where hurricanes coincided with COVID-19. Yet other studies in Greece and Mexico used historical data to examine how disaster responses might alter the post-disaster evolution of COVID-19 infections \cite{mavroulis2021geological, frausto2020covid}.

However, no research has explored the impacts of different vaccination rates on disaster evacuation operations. Here, we integrated different vaccine uptake levels into an age-structured epidemiological Susceptible-Exposed-Infectious-Recovered (SEIR) model to explore implications of alternative approaches to managing individuals exposed to COVID-19. Specifically, this study integrated real-world data, such as vaccine efficacy against infections \cite{kim2021looking, mahase2020covid, chagla2021adults}, daily vaccine uptake (\textit{i.e.}, the number of vaccines taken by evacuees per day) \cite{vaccine_kh}, evacuee population and age distributions \cite{populationdata}, and social contact matrices of every 5-year age group \cite{prem2017projecting}, into an age-structured SEIR model \cite{hogan2021within}. We aimed to quantitatively evaluate the efficacy of vaccine uptake and an emergency evacuation protocol that has been implemented in Taiwan, known as the Diversion protocol, in mitigating pandemic risks at an evacuation shelter. This Diversion protocol works as follows: for any mandatory evacuation during the COVID-19 pandemic, those who are in self-isolation due to COVID-like symptoms or in self-quarantine due to potential exposure to infections are diverted to non-congregate shelters (\textit{e.g.}, hotels) to prevent them from mingling with the uninfected general public at a typical congregate shelter like a community center or a large indoor arena \cite{taiwan}.

To more realistically model an evacuation during a pandemic, we chose Taoyuan and Jiaxian, two debris-flow-prone districts of Kaohsiung City, to study pre-debris-flow evacuation operations. In the early stages of the COVID-19 pandemic, Taiwan used big data analytics, strict border control, and proactive contact tracing and testing as the preemptive measures to successfully prevent what could have been the second highest number of confirmed cases in the world given the frequent flights and ships from China that arrive in Taiwan \cite{wang2020response, gardner_2020}. In this study, we also integrated COVID-19 data analytics, scenario simulations, and the Diversion disaster evacuation policy into our SEIR model simulations to investigate to what extent the Diversion protocol and different vaccine uptake levels can decrease the number of infections and delay the pandemic peak occurrences. Most importantly, we explored the implications of and insights into the dynamics of an evolving COVID-19 virus overlapping with the occurrence of debris flow evacuations. Our findings can help all levels of governments, NGOs, schools, private sectors, and citizens efficiently prepare for, respond to, and recover from the coupled impacts of a pandemic and a natural or humanitarian crisis.

\section*{Results}
\subsection*{Evacuation scenario analysis}
In our scenario simulation, we assumed that a debris flow evacuation was initiated in the Taoyuan and the Jiaxian Districts during the COVID-19 pandemic in September 2021. Since it would be challenging to know during a chaotic evacuation whether an individual evacuee had already been infected and was in the virus's latent period, we used age-stratified SEIR model simulations \cite{hogan2021within} to study to what extent separating the general public from those already in self-quarantine before the evacuation (following the Diversion protocol), combined with different vaccine uptake levels, can decrease the number of infections and slow down the outbreaks. The vaccines available in Taiwan in September 2021 were AstraZeneca (AZ) vaccine, with 70\% vaccine efficacy against infection \cite{chagla2021adults}, and the Moderna vaccine, with 95\% vaccine efficacy against infection \cite{mahase2020covid}. To investigate the impacts of different vaccine uptake levels on the pandemic evolution when evacuees stayed at a shelter for multiple days, we applied real-world data to infer the daily COVID-19 vaccine uptake starting 30 days prior to the mandatory evacuation order in each district (see Methods). In our scenario simulations of the limited vaccine uptake, 0.36\% of the evacuee population were vaccinated per day in the Taoyuan and Jiaxain districts. In the scenario simulations of the sufficient vaccine uptake, 1.86\% of the evacuee population were vaccinated daily.

It is important to note that in our evacuation scenario simulations, we considered only those who had been in self-quarantine before the evacuation and excluded those who had been in self-isolation. In Taiwan, people are notified if they were in contact with an infected person \cite{steinbrook2020contact, cheng2020contact}, and so we can assume that people in self-quarantine were in contact with someone with COVID-19 without knowing whether they were infected. In contrast, we can assume that people in isolation were sick and contagious. In our scenarios, since those who are in self-isolation would be known before an evacuation starts, they would be sent to a hospital or a non-congregate shelter like a quarantine hotel rather than to a congregate shelter. Therefore, we focused on those who were in self-quarantine prior to evacuation because they could be either Susceptible (uninfected) or Exposed (already infected but not yet infectious so still in the latent period). We did not assume that such individuals could be categorized as Infectious or Recovered because Infectious people are typically taken care of in a hospital, a government quarantine facility, or a quarantine hotel in Taiwan. Furthermore, we assumed Recovered people or those who were immune have immunity that lasts at least one year in light of the latest evidence from \cite{kang2022broad, guo2022sars, hogan2021within, vitale2021assessment} while some other latest studies found quicker re-infections \cite{sette2022immunological, dan2021immunological}. Situations where Recovered people become infected again were outside the scope of this study. Our goal was to explore to what extent the number of mild infections (IMild; no need for hospitalization) and severe infections (ICase; hospitalization required) \cite{hogan2021within} were decreased and the occurrences of pandemic peaks were delayed by different evacuation strategies in the Taoyuan and the Jiaxian districts.

\begin{figure}[H]\captionsetup{singlelinecheck = false, justification=justified}
	\centering
	\begin{subfigure}{0.9\textwidth} %
	    \caption{Mild cases (hospitalization not required)} %
		\includegraphics[width=\textwidth]{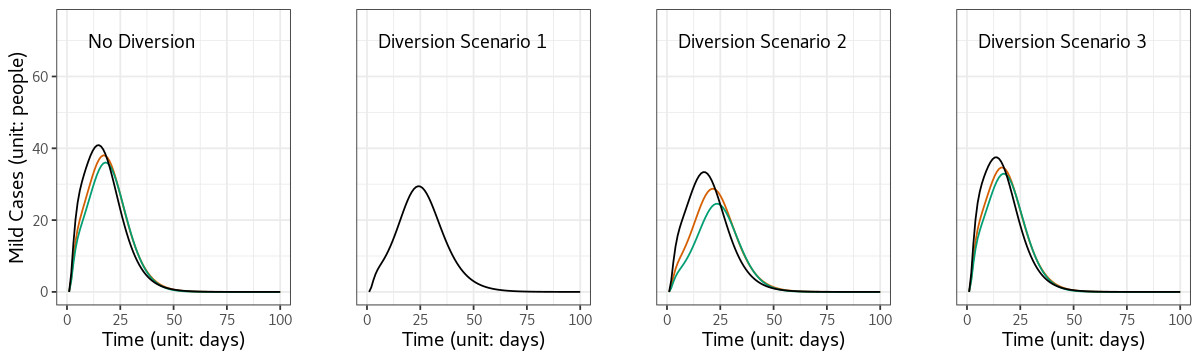}
	\end{subfigure}
	
	\begin{subfigure}{0.9\textwidth} %
        \caption{Severe cases (hospitalization required)} %
	    \includegraphics[width=\textwidth]{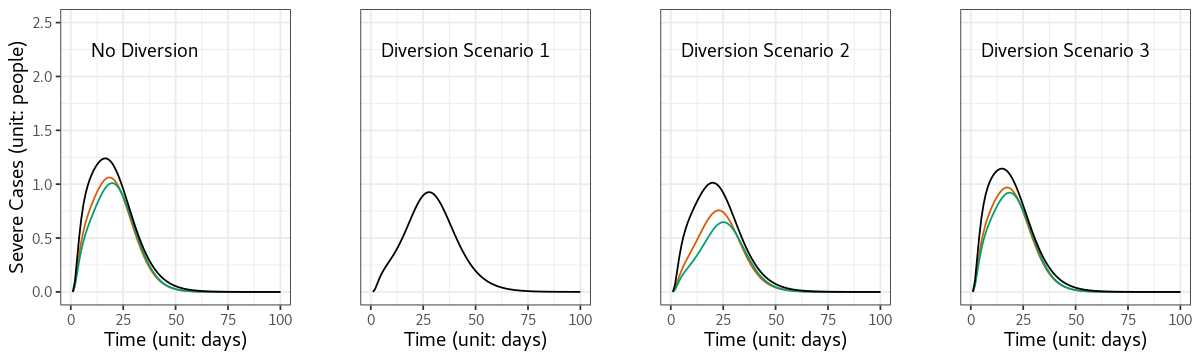}
	\end{subfigure}
	\caption{\textbf{Examples of the effects of various evacuation strategies on the number of mild and severe cases.} In these examples, there were 5 seeding cases in the Taoyuan District with $R_0$ = 2.5 and sufficient vaccine uptake since the first day of simulations at Stage 1, starting on September 1st, 2021, when there was no debris flow risk yet. The black curves represent the scenarios without vaccination. The orange curves mean that AstraZeneca (AZ) had been used, while the green curves signify that Moderna had been used since the first day of Stage 1 when there were no outbreaks yet but only a few exposed people in their latent period. The x-axis is the time of a pandemic emergence at the congregate shelter, which was assumed to start on September 30th, 2021 (Stage 2, when a debris flow evacuation was initiated). The y-axis is the number of mild or severe cases on each time step $t$ (day). In some extreme cases of our SEIR simulations, the number of Exposed people we assumed that we could divert was more than the original number of Exposed people among the general public by the time a disaster evacuation started. This implied that there were no seeding cases at the congregate shelter and no pandemic would occur. Therefore, the orange (AZ) and green (Moderna) curves in Diversion Scenario 1 were not shown. In this case, we assumed that the number of mild and severe cases stayed at 0 during the evacuees' stay at the congregate shelter. Moreover, to represent the situations where seeding cases were eliminated at the congregate shelter, and to quantify as well as compare the shift of the pandemic peak time using different evacuation strategies, we assumed a pandemic peak would occur on the 365th day. This assumption places the nominal possible outbreak outside of the 1-year time window we considered in this study. However, another outbreak might happen within one year due to the other sources of transmission outside the community, which is beyond the scope of this study.}\label{fig:imild_icase_example}  %
\end{figure}

We considered four possible cases in each district (Fig. \ref{fig:imild_icase_example}). The first was that everyone was evacuated together (No Diversion Scenario). The second case included the assumption that people in self-quarantine were evacuated towards non-congregate shelters to prevent them from mingling with the other uninfected but Susceptible general public at a congregate shelter, and all of those diverted people belonged to the Exposed category (Diversion Scenario 1). The third case was the same, but 50\% of them belonged to Susceptible and the other 50\% was Exposed (Diversion Scenario 2). For the fourth case, all of these self-quarantined people belonged to the Susceptible category (Diversion Scenario 3). Since the percentages of Susceptible and Exposed people among all evacuees were unknown during evacuation, we designed Diversion Scenarios 1, 2, and 3 to provide decision-makers with a comprehensive spectrum of possible percentages of Exposed people who could be diverted, ranging from the optimal (Diversion Scenarios 1) to the worst (Diversion Scenarios 3), when the Diversion protocol was enforced during a pandemic.

We also simulated the aforementioned four possible cases with different basic reproduction numbers ($R_0$ = 1.5, 2, 2.5, and 3). Seeding cases (5 and 10) served as the initial Exposed cases that would spread the virus and ignite the outbreaks after the latent period. We ran the SEIR simulations for two stages in each district. Stage 1 started on September 1st, 2021, and no debris flow risks were yet threatening. Stage 2 began on September 30th, 2021, and we assumed that a debris flow was about to occur so that a mandatory disaster evacuation was issued. We applied the number of Susceptible and Exposed on the 30th day of Stage 1 to the 1st day of Stage 2, and we ran another set of SEIR simulations for Stage 2 to examine to what extent the Diversion protocol and different vaccine uptake levels minimized the number of infections and delayed the occurrences of pandemic peaks while evacuees stayed at a congregate shelter.

Among all possible cases, the one producing the least number of mild and severe cases was Diversion Scenario 1 (Fig. \ref{fig:imild_icase_example}). Disaster managers were able to remove the maximum number of Exposed people (provided that 100\% of the people in self-quarantine belonged to Exposed) before they went to the congregate shelter and mingled with Susceptible individuals. In other words, disaster managers were able to decrease the maximum number of Exposed people who can spread the virus at the congregate shelter, and thus reduced the number of infections and delayed the occurrences of pandemic peaks (Fig. \ref{fig:imild_icase_example}).

When vaccine uptake was sufficient, vaccination decreased the number of infections and delayed the pandemic peak occurrences, especially when vaccination was combined with the Diversion protocol (Fig. \ref{fig:imild_icase_example}). When vaccine uptake was limited, the number of infections was close to the situations without vaccination (Supplementary information). This does not mean that vaccines cannot help combat the combined risks of a pandemic and a natural hazard. Rather, this is because in September 2021, the actual vaccine uptake was still quite limited in Taiwan (even though we used the maximum historical COVID-19 daily vaccine uptake in Kaohsiung City prior to September 1st, 2021 to calculate the proxy number of vaccines taken by evacuees per day) (see Methods). However, the Diversion protocol can still play an important role in decreasing infections and delaying the pandemic peak occurrences even when vaccine uptake is still insufficient (Supplementary information).

In terms of the number of infections and the occurrences of pandemic peaks, it was interesting to see that Diversion Scenario 3 and No Diversion Scenario were similar (Fig. \ref{fig:imild_icase_example}). Diversion Scenarios 1, 2, and 3 decreased the number of cases, and Scenario 1 delayed the peak of the pandemic by more days than Scenario 2. The best outcome was Diversion Scenario 1 (100\% of the diverted people belonged to Exposed), followed by Diversion Scenario 2 (50\% of the diverted people belonged to Exposed and 50\% of them belonged to Susceptible), and then Diversion Scenario 3 (100\% diverted people belonged to Susceptible). Although the pandemic peak occurred slightly sooner in Diversion Scenario 3, when compared to the No Diversion Scenario, the number of mild and severe cases was still fewer in Scenario 3 than the number in the No Diversion Scenario. Therefore, the No Diversion Scenario, wherein every evacuee mingled at the congregate shelter, had the worst outcome. Even though policy makers cannot actively choose which scenario they want and make it happen, the Diversion protocol is still valuable in situations where the ability to divert Exposed people is not optimal (Scenarios 2 and 3). In the following sections, we compare various evacuation scenarios through topical and in-depth analysis. We include all of the SEIR scenario simulation results in the tables in Supplementary information.

\subsection*{The number of infections}
Considering Diversion scenarios 1, 2, and 3 with 5 and 10 seeding cases, without any vaccination, and with a reproduction number between 1.5 and 3, the Diversion disaster evacuation protocol on average decreased the number of mild infections by 38.15\% and 52.55\% and the number of severe cases by 37.79\% and 52.27\% when the outbreak peaked in, respectively, the Taoyuan District and the Jiaxian District compared to those situations where everyone went to the congregate shelter without the Diversion protocol. Meanwhile, under such conditions and the evacuation protocol, limited vaccination can on average slightly decrease the number of positive cases---by 39.42\% and 53.28\% for mild infections and by 39.68\% and 53.45\% for severe cases when the pandemic peaked in, respectively, the Taoyuan District and the Jiaxian District. With sufficient vaccine uptake and the Diversion protocol, the number of mild cases can on average be significantly decreased by 49.84\% and 60.12\%, while the number of severe cases can be reduced by 51.24\% and 61.38\% in these two districts, respectively. Both sufficient vaccine uptake and the Diversion protocol were crucial in mitigating the number of infections at a congregate shelter.

In general, the more seeding cases at the beginning of evacuations, the more mild and severe infections (IMild and ICase) occurred after people stayed at the congregate shelter for several days. However, after evacuations of people to the congregate shelter, fewer seeding cases (5) resulted in higher numbers of mild and severe infections for evacuees during Stage 2. This occurred when $R_0$ was larger than 2 in Stage 1 but this effect faded when $R_0$ decreased. This happened in our case when $R_0$ decreased from 3 to 1.5. For example, when $R_0$ was 3 and the seeding cases increased from 5 to 10 in Stage 1, the Exposed population decreased by 13.11\%, 11.97\%, and 10.62\% for the scenarios of, respectively, no vaccination, AZ, and Moderna in the Taoyuan District on the 30th day of Stage 1. In contrast, when $R_0$ decreased from 3 to 1.5 under the same condition, the Exposed population increased by 61.54\%, 72.73\%, and 72.73\% for the scenarios of no vaccination, AZ, and Moderna, respectively.

\begin{figure}[H]\captionsetup{singlelinecheck = false, justification=justified}
	\centering
	\begin{subfigure}{0.49\textwidth} %
	    \caption{Seeding cases = 5} %
		\includegraphics[width=\textwidth]{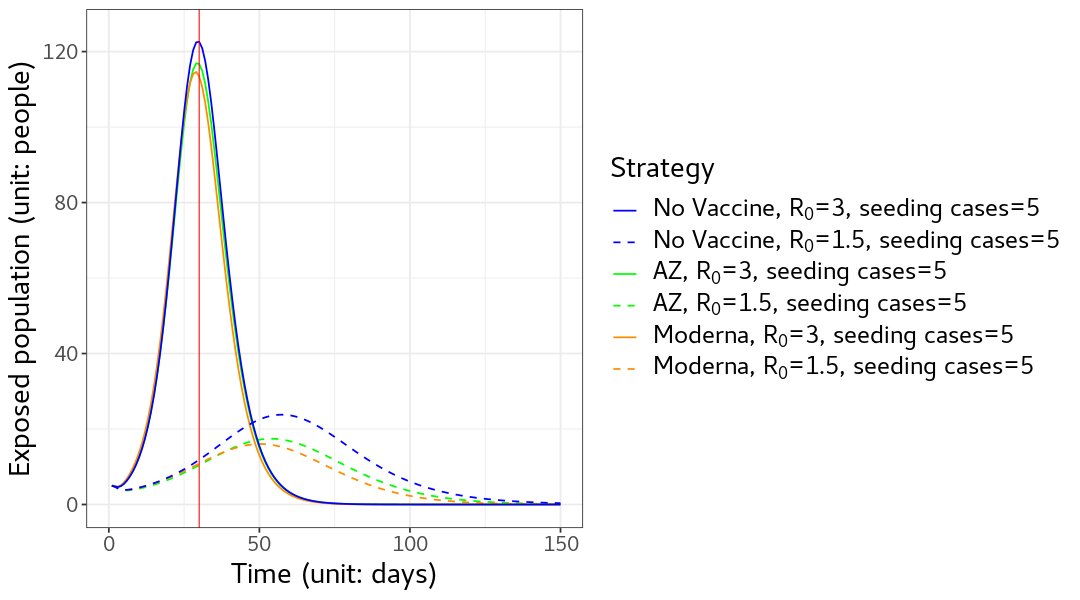}
	\end{subfigure}
	\vspace{0.1cm} %
	\begin{subfigure}{0.49\textwidth} %
	    \caption{Seeding cases = 10} %
		\includegraphics[width=\textwidth]{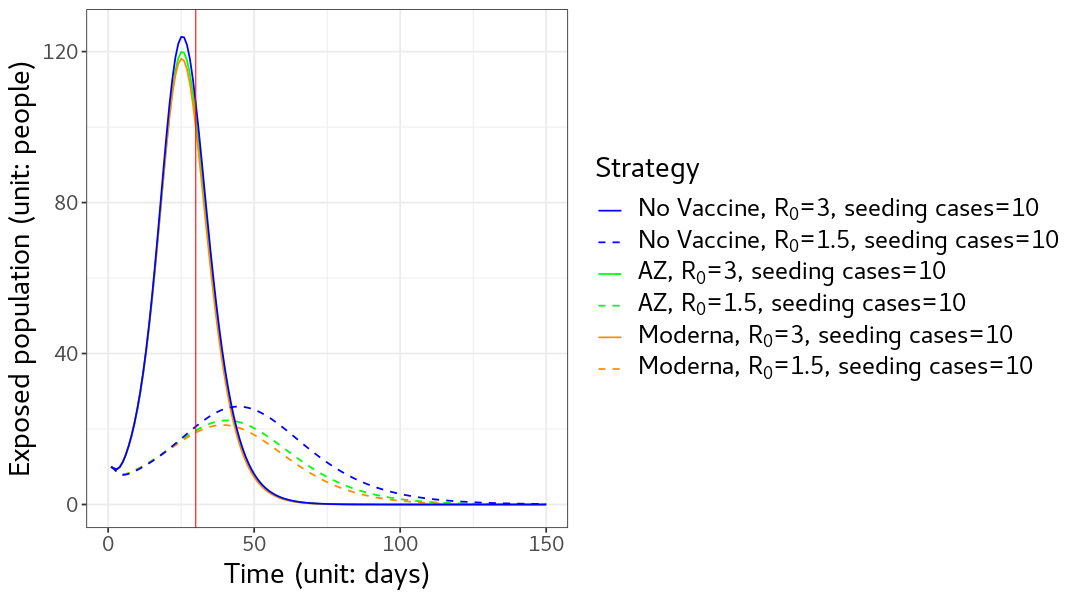}
	\end{subfigure}
	\caption{\textbf{An example of a comparison between the number of Exposed population with 5 and 10 seeding cases prior to the evacuations (Stage 1).} Here we used the Taoyuan District to illustrate the impacts of different seeding cases (Exposed population) in Stage 1 on the pandemic evolution after the disaster evacuation began (Stage 2). The red vertical line indicates that the debris flow evacuation occurred on Day 30 in Stage 1, which started on September 1st, 2021, and there was not yet any debris flow risk in Stage 1. The solid curves refer to the scenarios with the $R_0$ of 3, while the dashed curves represent scenarios with the $R_0$ of 1.5. The x-axis is the time series of a pandemic emergence of Stage 1. The y-axis is the number of Exposed people on each time step $t$ (day).}\label{fig:stage_1_exposed}.  
\end{figure}

We also found that more seeding cases (10) led to an earlier occurrence of the peak of the Exposed population in Stage 1 (Fig. \ref{fig:stage_1_exposed}b). In other words, more seeding cases (10) would cause the curve of Exposed population to rise and fall more quickly than fewer seeding cases (5) in Stage 1 (Fig. \ref{fig:stage_1_exposed}). In this case, when the mandatory disaster evacuation occurred on the 30th day of the pandemic (the red vertical line in Fig. \ref{fig:stage_1_exposed}), the number of Exposed population with 10 seeding cases (the intersection of the solid curves and the red vertical line in Fig. \ref{fig:stage_1_exposed}b) was less than the one with 5 seeding cases (the intersection of the solid curves and the red vertical line in Fig. \ref{fig:stage_1_exposed}a) in Stage 1. In Fig. \ref{fig:stage_1_exposed}a, the peak for the $R_0$ of 3 occurred at Day 30 whilst the $R_0$ of 1.5 peaked after Day 50. In Fig. \ref{fig:stage_1_exposed}b, the peak for the $R_0$ of 3 occurred around Day 25 prior to the evacuation but the $R_0$ of 1.5 peaked earlier around Day 45.

\subsection*{The occurrence of pandemic peaks}
Implementing the Diversion evacuation strategy decreased the number of mild and severe infections for both groups (Susceptible or Exposed) in self-quarantine. However, if people who were diverted to non-congregate shelters in compliance with the Diversion protocol all belonged to the Susceptible group, the pandemic peaks would arrive slightly earlier (Diversion Scenario 3 in Fig. \ref{fig:imild_icase_example}). Only when some of these people belonged to the Exposed group and were diverted to non-congregate shelters would the pandemic peaks occur later.

As mentioned earlier, if the number of Exposed evacuees we assume that we could divert was greater than the original number of Exposed evacuees at the start of the evacuation, the outbreak risk was eliminated at the congregate shelter, and we assumed the peak would be delayed until the 365th day of Stage 2, which places the nominal possible outbreak outside of the 1-year time window we considered in this study. In other words, in this study, if a peak is delayed until the 365th day since evacuation, this means that the pandemic risk has been totally eliminated under that condition. If disaster managers implemented the Diversion protocol without any vaccination for each basic reproduction number (1.5, 2, 2.5, or 3) and each seeding case (5 or 10), the peak of mild infections, on average, can be delayed by 84.83 and 143.42 more days and the peak of severe infections can be delayed by 84.71 and 141.42 more days in the Taoyuan District and the Jiaxian District, respectively.

With limited vaccine uptake, the peak of infections occurred slightly later in most scenarios. On average, in terms of mild infections, implementing the Diversion protocol with limited vaccine uptake delayed the pandemic peaks by 86 and 143.67 more days in the Taoyuan District and the Jiaxian District, respectively. As for the severe cases, the peak was delayed, on average, by 85.6 and 142.19 more days, respectively, in these two districts. With sufficient vaccine uptake and the Diversion protocol, disaster managers can delay the pandemic peaks to much later days---by 122.06 and 172.6 more days for the mild infection peaks and by 122.17 and 171.52 more days for the severe infection peaks, respectively, in these two districts. In other words, the combined effect of sufficient vaccine uptake and the Diversion protocol can significantly delay pandemic peaks. In sum, if pandemic peaks can be delayed, evacuees have less exposure to virus transmission in a congregate shelter.

\subsection*{The impact of seeding cases on the pandemic evolution}

\begin{figure}[H]\captionsetup{singlelinecheck = false, justification=justified}
	\centering
	\begin{subfigure}{0.49\textwidth} %
		\caption{Peak timing of mild cases under limited vaccine uptake} %
		\includegraphics[width=\textwidth]{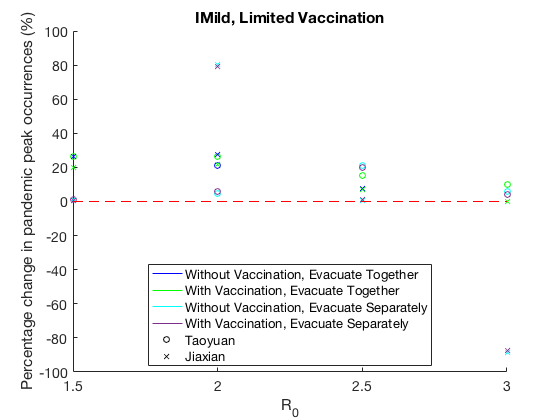}
	\end{subfigure}
    \vspace{0.1 cm} %
	\begin{subfigure}{0.49\textwidth} %
		\caption{Peak timing of severe cases under limited vaccine uptake} %
		\includegraphics[width=\textwidth]{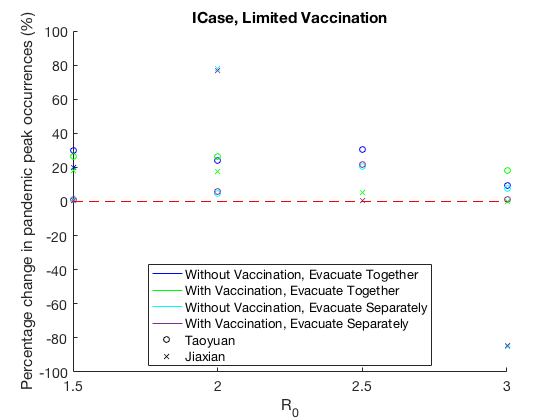}
	\end{subfigure}
	\vspace{0.1 cm} %
	\begin{subfigure}{0.49\textwidth} %
		\caption{Peak timing of mild cases under sufficient vaccine uptake} %
		\includegraphics[width=\textwidth]{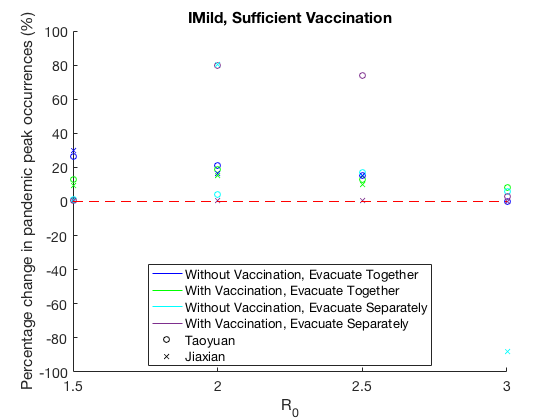}
	\end{subfigure}
    \vspace{0.1 cm} %
	\begin{subfigure}{0.49\textwidth} %
		\caption{Peak timing of severe cases under sufficient vaccine uptake} %
		\includegraphics[width=\textwidth]{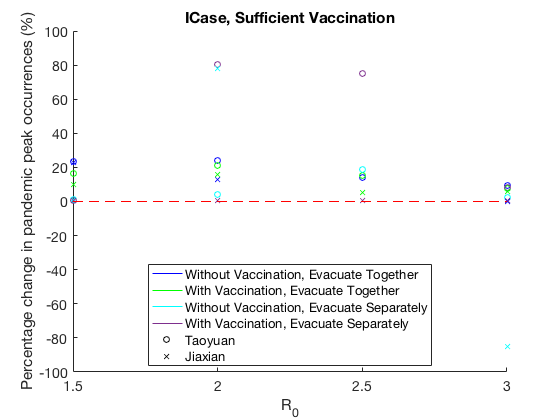}
	\end{subfigure}
	
	\caption{\textbf{The effect of different seeding cases on the pandemic peak occurrences.} Circle points (Taoyuan District). Cross points (Jiaxian District). Points above the red dashed line signify that the pandemic peak occurrences occurred later in the scenarios with 5 seeding cases, compared to those with 10 seeding cases, and vice versa. The x-axis is the basic reproduction number $R_0$ and the y-axis is the percentage change in the peak occurrences of mild cases (Fig. \ref{fig:seeding_10_5_peak_time}a, c) and severe cases (Fig. \ref{fig:seeding_10_5_peak_time}b, d), calculated as equation (\ref{eq:peak_time_percentage_change_seeding}).}\label{fig:seeding_10_5_peak_time} 
\end{figure}

\begin{figure}[H]\captionsetup{singlelinecheck = false, justification=justified}
	\centering
	\begin{subfigure}{0.49\textwidth} %
		\caption{Mild cases at the peak under limited vaccine uptake} %
		\includegraphics[width=\textwidth]{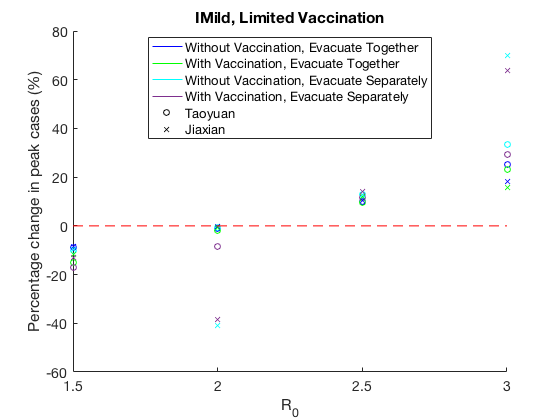}
	\end{subfigure}
	\vspace{0.1 cm} %
	\begin{subfigure}{0.49\textwidth} %
		\caption{Severe cases at the peak under limited vaccine uptake} %
		\includegraphics[width=\textwidth]{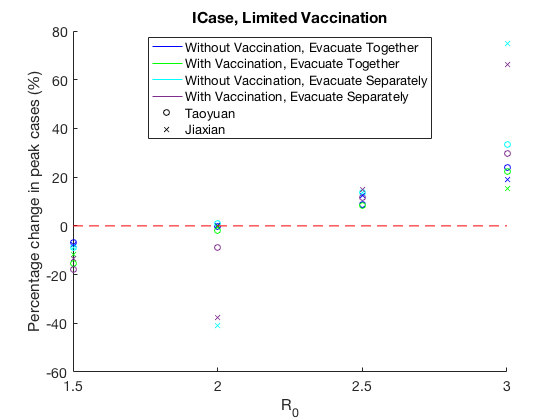}
	\end{subfigure}
	
	\vspace{0.1 cm} %
	\begin{subfigure}{0.49\textwidth} %
		\caption{Mild cases at the peak under sufficient vaccine uptake} %
		\includegraphics[width=\textwidth]{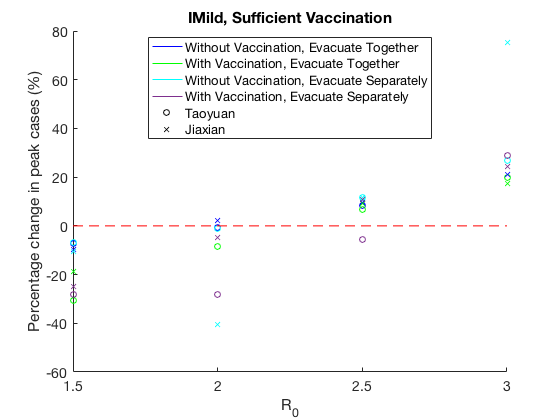}
	\end{subfigure}
    \vspace{0.1 cm} %
	\begin{subfigure}{0.49\textwidth} %
		\caption{Severe cases at the peak under sufficient vaccine uptake} %
		\includegraphics[width=\textwidth]{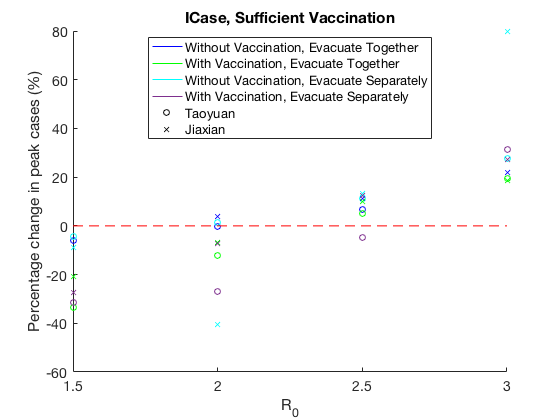}
	\end{subfigure}
	
	\caption{\textbf{The effect of different seeding cases on the number of infections at the pandemic peaks.} Circle points (Taoyuan District). Cross points (Jiaxian District). Points above the red dashed line signify that there were more infections in the scenarios with 5 seeding cases, compared to those with 10 seeding cases, and vice versa. The x-axis is the basic reproduction number $R_0$ and the y-axis is the percentage changes in the number of mild cases (Fig. \ref{fig:seeding_10_5_peak_cases}a, c) and severe cases (Fig. \ref{fig:seeding_10_5_peak_cases}b, d), calculated as equation (\ref{eq:peak_cases_percentage_change_seeding}).}\label{fig:seeding_10_5_peak_cases} %
\end{figure}

In both the Taoyuan District and the Jiaxian District, decreasing seeding cases from 10 to 5 delayed the pandemic peak occurrences in most situations, regardless of the kinds of disaster evacuation strategies implemented (Fig. \ref{fig:seeding_10_5_peak_time}). Some exceptions may occur due to the different vaccine uptake levels and the number of seeding cases by the time an evacuation is issued (Fig. \ref{fig:stage_1_exposed}). 
As for the impacts of decreasing seeding cases on the number of infections at the pandemic peaks, the consequences depended on the basic reproduction number $R_0$. Under the conditions of smaller $R_0$ ($R_0 = 1.5\ or\ 2$), decreasing the seeding cases from 10 to 5 helped decrease up to 40\% of the number of mild cases and severe cases (Fig. \ref{fig:seeding_10_5_peak_cases}a-d). However, with larger $R_0$ ($R_0 = 2.5\ or\ 3$), decreasing the seeding cases from 10 to 5 led to more cases. Regardless of whether the vaccine uptake was limited or sufficient, these special situations occurred because more seeding cases (10) would cause the curve of Exposed population to rise and fall more quickly than fewer seeding cases (5) in Stage 1, consistent with Fig. \ref{fig:stage_1_exposed}.

\begin{equation} \label{eq:peak_time_percentage_change_seeding}{Percentage\ Change\ in\ the\ Pandemic\ Peak\ Occurrences\ (\%)}= \frac{Time_{seeding=5}-Time_{seeding=10}}{Time_{seeding=10}}\times 100 \\
\end{equation}

\begin{equation} \label{eq:peak_cases_percentage_change_seeding}{Percentage\ Change\ in\ the\ Number\ of\ Cases\ (\%)}= \frac{Cases_{seeding=5}-Cases_{seeding=10}}{Cases_{seeding=10}}\times 100 \\
\end{equation}

\subsection*{The impact of evacuees’ age distribution on the pandemic evolution}

\begin{figure}[H]\captionsetup{singlelinecheck = false, justification=justified}
	\centering
	\begin{subfigure}{0.49\textwidth} %
		\caption{Peak timing of mild cases under limited vaccine uptake} %
		\includegraphics[width=\textwidth]{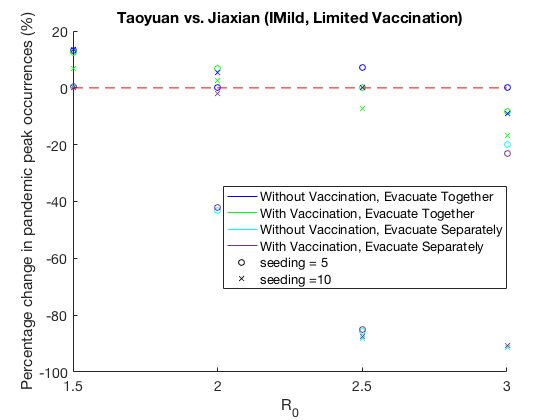}
	\end{subfigure}
	\vspace{0.1 cm} %
	\begin{subfigure}{0.49\textwidth} %
		\caption{Peak timing of severe cases under limited vaccine uptake} %
		\includegraphics[width=\textwidth]{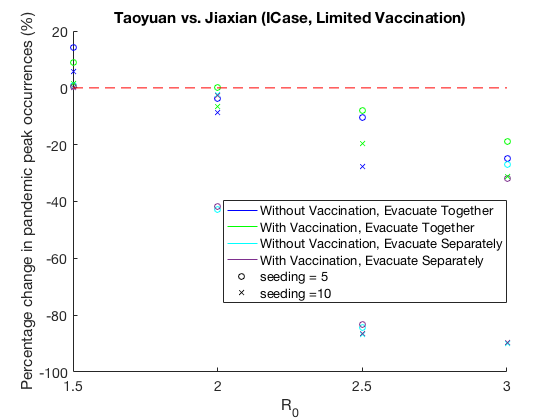}
	\end{subfigure}
	
	\centering
	\begin{subfigure}{0.49\textwidth} %
		\caption{Peak timing of mild cases under sufficient vaccine uptake} %
		\includegraphics[width=\textwidth]{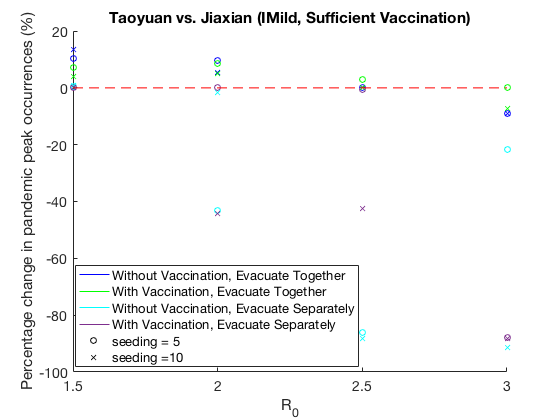}
	\end{subfigure}
	\vspace{0.1 cm} %
	\begin{subfigure}{0.49\textwidth} %
		\caption{Peak timing of severe cases under sufficient vaccine uptake} %
		\includegraphics[width=\textwidth]{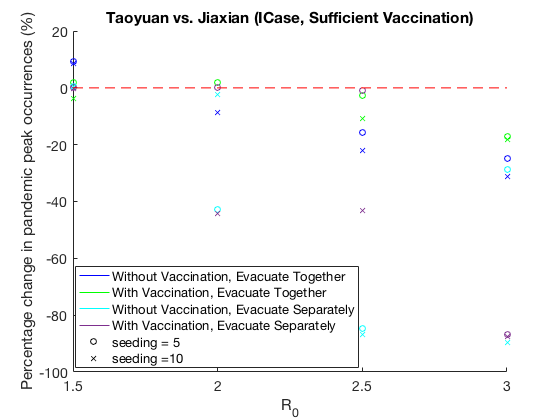}
	\end{subfigure}

	\caption{\textbf{The comparison of the pandemic peak occurrences between the Taoyuan District and the Jiaxian District with different evacuation strategies and vaccine uptake levels.} Circle points (5 seeding cases). Cross points (10 seeding cases). Points below the red dashed line means that the pandemic peaks occurred sooner in the Taoyuan District, compared to the Jiaxian District, and vice versa. The x-axis is the basic reproduction number $R_0$, and the y-axis is the percentage change in the peak occurrences of mild cases (Fig. \ref{fig:TJ_peak_time}a, c) and severe cases (Fig. \ref{fig:TJ_peak_time}b, d), calculated as equation (\ref{eq:peak_time_percentage_change}).}\label{fig:TJ_peak_time} %
\end{figure}

With a group of evacuees that consisted primarily of a young adult population of around 25--40 years old in the Taoyuan District (Supplementary Fig. S1), the pandemic peaks tended to occur sooner (Fig. \ref{fig:TJ_peak_time}), compared to a majority elderly group, such as in the Jiaxian District (Supplementary Fig. S1). This is because young and middle-aged people have more contacts within their own age groups and with other age groups in the age-stratified mixing social contact matrices \cite{hogan2021within, prem2017projecting} compared to elderly people.

More specifically, when $R_0$ was larger, the effect of the young adult population as the majority in a disaster-prone area on making the pandemic peak occurrences occur sooner was more explicit (Fig. \ref{fig:TJ_peak_time}). When $R_0$ was 3, the peak of both mild and severe cases occurred more quickly after the disaster evacuation was initiated in the Taoyuan District, regardless of whether vaccination or the Diversion protocol was implemented or not. When $R_0$ was 2 or 2.5, the peak of severe cases occurred earlier in all kinds of scenarios in the Taoyuan District than in the Jiaxian District (Fig. \ref{fig:TJ_peak_time}b, d), while the one of mild cases occurred later in some of the scenarios, especially those without the Diversion protocol (Fig. \ref{fig:TJ_peak_time}a, c). Although the pandemic peaks occurred later in the Taoyuan District under the condition of $R_0 = 1.5$, the overall trend was that the young adult evacuee population led to earlier pandemic peak occurrences under the conditions of larger $R_0$. The pattern of the percentage change in pandemic peak occurrences was similar across situations with limited and sufficient vaccine uptake levels among evacuees (Fig. \ref{fig:TJ_peak_time}).

\begin{equation} \label{eq:peak_time_percentage_change}{Percentage\ Change\ in\ the\ Pandemic\ Peak\ Occurrences\ (\%)}= \frac{Time_{Taoyuan}-Time_{Jiaxian}}{Time_{Jiaxian}}\times 100 \\
\end{equation}

\begin{figure}[H]\captionsetup{singlelinecheck = false, justification=justified}
	\centering
	\begin{subfigure}{0.49\textwidth} %
		\caption{Mild cases at the peak under limited vaccine uptake} %
		\includegraphics[width=\textwidth]{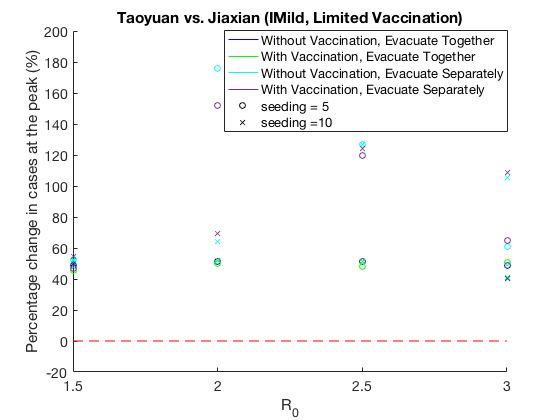}
	\end{subfigure}
	\vspace{0.1 cm} %
	\begin{subfigure}{0.49\textwidth} %
		\caption{Severe cases at the peak under limited vaccine uptake} %
		\includegraphics[width=\textwidth]{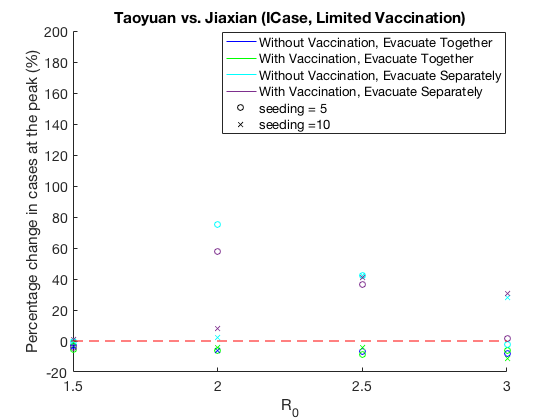}
	\end{subfigure}
	
	\centering
	\begin{subfigure}{0.49\textwidth} %
		\caption{Mild cases at the peak under sufficient vaccine uptake} %
		\includegraphics[width=\textwidth]{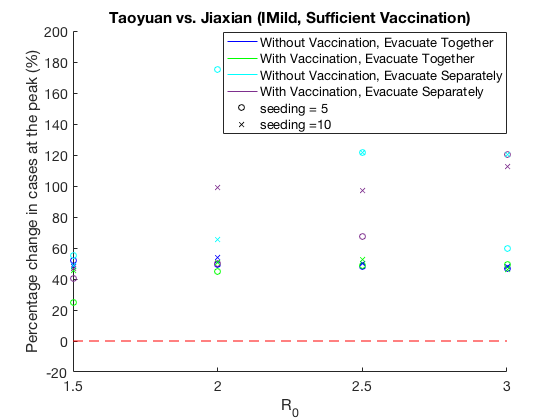}
	\end{subfigure}
	\vspace{0.1 cm} %
	\begin{subfigure}{0.49\textwidth} %
		\caption{Severe cases at the peak under sufficient vaccine uptake} %
		\includegraphics[width=\textwidth]{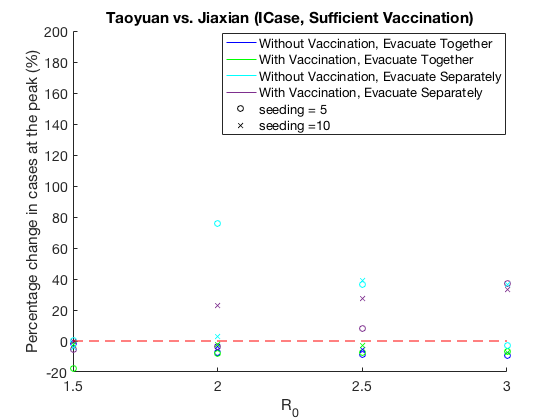}
	\end{subfigure}

	\caption{\textbf{The comparison of the number of infections at the pandemic peaks between the Taoyuan District and the Jiaxian District with different evacuation strategies and vaccine uptake levels.} Circle points (5 seeding cases). Cross points (10 seeding cases). Points above the red dashed line indicate that there are more infections in the Taoyuan District than the Jiaxian District, and vice versa. The x-axis is the basic reproduction number $R_0$ and the y-axis is the percentage change in the number of mild cases (Fig. \ref{fig:TJ_peak_cases}a, c) and severe cases (Fig. \ref{fig:TJ_peak_cases}b, d), calculated as equation (\ref{eq:peak_cases_percentage_change}).}\label{fig:TJ_peak_cases} 
\end{figure}

Similar to the trend in the pandemic peak occurrences, due to the higher transmission probability from the strong assortativity of social contacts in young adult population, there were more infections in the Taoyuan District in most of the situations (Fig. \ref{fig:TJ_peak_cases}). In particular, regardless of which disaster evacuation strategy was implemented, there were more mild infections in all of the scenarios in the Taoyuan District than in the Jiaxian District (Fig. \ref{fig:TJ_peak_cases}a, c). However, when $R_0$ was smaller ($R_0$ = 1.5), or the Diversion protocol was not implemented, a community with the elderly as the majority, like the Jiaxian District, tended to suffer from more severe cases, compared with the Taoyuan District, regardless of whether vaccination was implemented or not (Fig. \ref{fig:TJ_peak_cases}b, d). Implementing the Diversion protocol could potentially have mitigated the consequences of exposing the vulnerable elderly to the higher risks of severe infections when $R_0$ was larger ($R_0$ = 2, 2.5, or 3). Regardless of whether vaccine uptake was limited or sufficient, the relationship between $R_0$ and the percentage change in the number of infections at the pandemic peaks was almost the same (compare Fig. \ref{fig:TJ_peak_cases}a, b with Fig. \ref{fig:TJ_peak_cases}c, d).

\begin{equation} \label{eq:peak_cases_percentage_change}{Percentage\ Change\ in\ the\ Number\ of\ Positive\ Cases\ (\%)}= \frac{Cases_{Taoyuan}-Cases_{Jiaxian}}{Cases_{Jiaxian}}\times 100 \\
\end{equation}

\section*{Discussion}
Our study found that diverting those who were in self-quarantine, due to potential exposure, towards non-congregate shelters to prevent them from mingling with others at a congregate shelter (the Diversion disaster evacuation protocol) not only decreased the number of infections, but also delayed the occurrences of pandemic peaks. Enforcing both sufficient vaccine uptake and the Diversion protocol can significantly reduce the maximum number of infections and delay the pandemic peaks. Even if vaccine uptake is limited, the Diversion protocol plays a crucial role in mitigating outbreak risks when evacuees stay at a shelter for several days.

The best possible outcome is when a disaster manager separated the maximum number of Exposed people from members of the general public (Diversion Scenario 1). The number of mild and severe cases was minimized, and the peak occurrence was delayed to the latest day, compared to all other scenarios (Fig. \ref{fig:imild_icase_example}). The second-best outcome was Diversion Scenario 2, where half the removed people belonged to Susceptible, while the other half belonged to Exposed. When the number of infections peaked in Diversion Scenario 3, the number of infections was still rising in No Diversion Scenario. Even though the peak occurred sooner in Diversion Scenario 3 compared to the No Diversion Scenario, the No Diversion Scenario was still the worst possible outcome. Although the optimal Scenario 1 might be infeasible in real-world disaster evacuations due to the challenges of identifying COVID-19 cases in disaster evacuation settings, the Diversion protocol is still helpful in situations where the ability to divert Exposed people is not 100\% (Diversion Scenarios 2 and 3).

In situations without vaccination, across all conditions with various seeding cases and basic reproduction numbers, the Diversion protocol significantly decreased the number of infections, on average, by 45.18\% and delayed the pandemic peaks by 113.6 more days, compared with the No Diversion Scenario in the Taoyuan District and the Jiaxian District. If the Diversion protocol was not implemented, rolling out limited vaccination and sufficient vaccination decreased the number of infections, on average, by 21.9\% and 94.47\%, respectively, while bringing forward the pandemic peak by 1.38 and 6.19 days, respectively, compared with scenarios without vaccination. Although the vaccination on average brought forward the pandemic peak, rolling out limited vaccination and sufficient vaccination delayed the pandemic peak, on average, by 0.94 and 5.13 more days when $R_0$ was larger ($R_0$ = 2, 2.5, or 3).

In terms of reducing infections and delaying the pandemic peak occurrences, the combined effects of vaccination and the Diversion protocol were superior to implementing only one or the other. Notably, the Diversion protocol was crucial in mitigating infections and delaying the peak occurrences when vaccines are limited. With the Diversion protocol, we found that the limited vaccine uptake can deduct only an additional 1.27\% of the number of infections and delay the pandemic peaks by only 0.77 more days compared to the situations without any vaccination. However, limited vaccine uptake can still make a difference in preventing severe cases and running out of beds, ventilators, and other medical resources. Sufficient vaccine uptake can further keep the healthcare systems from the verge of collapse because it can reduce an additional 10.48\% of the number of infections and delay pandemic peaks by 33.47 more days compared to the situations without any vaccination. We can thus interpret the role of vaccination in two ways: (1) If vaccines are relatively limited, it is imperative to implement the Diversion protocol to mitigate the risks of cascading COVID-19 outbreaks at a congregate shelter. (2) If vaccines are sufficient, evacuees who live in a disaster-prone region must take the vaccines as soon as possible, and the Diversion protocol should be implemented, so that more infections can be avoided, and the pandemic peaks can be delayed at a congregate shelter compared to scenarios without such strategies.

We found that more seeding cases at Stage 1 did not necessarily lead to more infections at Stage 2 (Fig. \ref{fig:stage_1_exposed}). $R_0$ and at which moment a debris flow evacuation was initiated during the pandemic mattered. In this study, we assumed that a debris flow evacuation was issued on the 30th day of Stage 1, which was the beginning of Stage 2. With a larger basic reproduction number ($R_0 = 2.5\ or\ 3$), the situations with more seeding (10) caused the curve of Exposed population to rise and fall more quickly at Stage 1 (Fig. \ref{fig:stage_1_exposed}). Therefore, the number of Exposed people at the beginning of Stage 2 was less than the situations with fewer seeding cases (5) because the number of Exposed people had peaked earlier and then dropped to a lower level by the time when an evacuation was initiated (Fig. \ref{fig:stage_1_exposed}). However, under the condition of a smaller basic reproduction number ($R_0 = 1.5\ or\ 2$), more seeding cases at Stage 1 led to more Exposed people at the beginning of Stage 2.

In terms of the pandemic peak occurrences, only when at least one of the diverted evacuees who were in self-quarantine belonged to the Exposed population (Diversion Scenarios 1 and 2) would the peaks of the pandemics be delayed with the implementation of the Diversion protocol. The evacuees' age distribution affected the evolution of a pandemic. A group of evacuees with young adult population as the majority (like the Taoyuan District) tended to have the pandemic peak occurrences earlier than the majority elderly group (like the Jiaxian District) (Fig. \ref{fig:TJ_peak_time}). The more serious a pandemic, the more explicit this phenomenon. Moreover, the young adult evacuee group generally led to more mild and severe cases, compared with the older-age one. The exception was that when the Diversion protocol was not implemented, the one with the elderly as the major population suffered from more severe infections (Fig. \ref{fig:TJ_peak_cases}) implying that the elderly with higher health risks were vulnerable to severe symptoms and would require more medical resources. The good news is that implementing the Diversion disaster evacuation protocol can mitigate this impact on the elderly, who were more vulnerable to COVID-19 than the young adult group.

This study might be more relevant to the evacuation situations where evacuees got vaccinated not long before an initiation of disaster evacuation because we assumed the vaccine effectiveness would stay the same within the time window we considered in this study. However, given the fact that the protection from vaccines would diminish over time \cite{centers2022centers, chemaitelly2022waning, piechotta2022waning}, it is worthwhile considering the waning of COVID-19 vaccine effectiveness in future research. This study used the daily vaccine uptake to investigate how increasing vaccine coverage among evacuees would dynamically influence compound disaster evacuation. Instead of using daily vaccine uptake, we could also use static conditioning factors, such as the fraction of evacuees vaccinated at the start of the evacuation, to study how the pandemic emerges and whether it is worth implementing the Diversion protocol based on cost-benefit analysis. Future work investigating to what extent other preemptive measures, such as rapid antigen testing, mitigates the virus transmission in a congregate shelter during a pandemic will help disaster managers and public health officers understand how to coordinate the medical and disaster relief resources.

To effectively evacuate people before any disasters or humanitarian crises occur, governments should establish a special needs registry to track those in self-quarantine and self-isolation, especially those who may need additional assistance during the evacuations, such as the elderly, pregnant women, children, and people with disabilities and high health risks. Our study demonstrated that the Diversion disaster evacuation protocol can be crucial in mitigating infections and delaying pandemic peaks at a congregate shelter even if vaccines are limited, such as in developing countries. The combination of the Diversion protocol and sufficient vaccine uptake can mitigate the most compound disaster risks by delaying the pandemic peak occurrences and reducing infections when a natural disaster or a humanitarian crisis strikes during a pandemic.

Evacuees sent to congregate shelters may be concerned that they will be exposed to higher pandemic risks and receive different and unequal treatment due to a large group of people in a poorly ventilated indoor space, compared to those who are diverted to non-congregate shelters. Therefore, it is crucial to ensure that preemptive public health measures should be enforced at the congregate shelter for the evacuees, such as cleaning and disinfecting the shelter and providing medical support or COVID-19 antigen rapid tests. For those who are diverted to non-congregate shelters, they may not want to be separated from their family members and pets or the rest of the evacuees. So, emotional support from their families and pets, as well as mental wellness counseling from governments, should be provided for them in addition to health monitoring and tracking.

When it becomes harder and uneconomical to maintain virus transmission as close to zero as possible due to the rapid spread of more contagious SARS-CoV-2 variants, many countries are shifting from the ``Zero-COVID'' policy towards the ``Coexisting with the Virus'' strategy \cite{normile2022china, normile2021zero}. It is imperative to mitigate pandemic risks by minimizing the severe cases and effectively managing mild cases to avoid overwhelming the capacity of healthcare resources and systems, especially during a disaster evacuation. Similar to the ``Coexisting with the Virus'' strategy, we need to consider the ``Coexisting with the Disaster'' strategy when it becomes more challenging than ever to prevent compound disasters from occurring due to global climate change. This study demonstrated and quantified the effectiveness of the Diversion disaster evacuation protocol and different vaccine uptake levels in mitigating virus transmission during a pandemic, which can help realize the ``Coexisting with the Disaster'' and the ``Coexisting with the Virus'' strategies. Most importantly, this study aims to better inform our society that successful disaster management and the window of opportunity for effective compound disaster mitigation, preparedness, response, and recovery require not only all-hazard crisis response and evacuation plans that consider the cascading effects of a pandemic and any possible natural hazards or humanitarian crises, but also the community-engaged participation from all levels and fields of policy-makers, CDC officials, disaster managers, school leadership, and citizens.

\section*{Methods}
\subsection*{Dataset}
We applied an age-structured SEIR-type mechanistic epidemiological model of COVID-19 dynamics \cite{hogan2021within} to two districts. Moreover, we conducted quantitative computation by integrating various evacuation scenarios into the SEIR simulations. There were 736 people and 601 people on the debris flow evacuation registry in the Taoyuan District and the Jiaxian District, respectively \cite{jan_zeng_2021}. By the time an evacuation was initiated at Stage 2 in this study (the 30th day of Stage 1 since the pandemic started), some of those people became Infectious and Recovered. Infectious people were known before the evacuation, so they were sent to hospitals or non-congregate shelters rather than at the congregate shelter with members of the general public. We assume that the immunity of those Recovered people lasts at least one year, so we are not concerned about them in the scope of this study. Those who were eventually evacuated to the congregate shelter were less than the original number of evacuees on the registry. Given that a typical congregate shelter in the Taoyuan District and the Jiaxian District can accommodate hundreds of evacuees \cite{jan_zeng_2021}, the number of evacuees who stayed at the congregate shelter we considered in this study was reasonable.

In addition, a congregate shelter can accommodate evacuees for 14 days to 6 months in short-term and mid-term evacuation planning and sheltering in Taiwan \cite{disaster_tw}. If the disaster recovery takes longer than 6 months to rebuild the community, evacuees will be transferred to long-term shelters. In this study, all pandemics peaked around 10 days--1 month and finished (curves became flattened) around 1--3 months even though the SEIR model we applied in this study simulates what would happen in the upcoming one year. Therefore, the time windows of the pandemic in this study were within a reasonable length of time for short-term and mid-term evacuation planning and sheltering.

In this study, we focused on Susceptible and Exposed people, who might cause an outbreak when they mingled at the congregate shelter. The idea of the Diversion protocol was to decrease the number of Exposed people at the congregate shelter to mitigate the pandemic risks. However, by the time an evacuation was initiated, disaster managers would not be able to identify the Susceptible or Exposed. So, the Diversion protocol focused on those who were in self-quarantine due to potential exposures to infections because they were more likely to be Exposed people in the latent period, compared to other general public. Based on the data published by the Kaohsiung government in September, 2021 \cite{covid19data}, on average, there were $N = 66$ people in self-quarantine per district who could be either Susceptible or Exposed, which, again, was unknown to disaster managers. The Diversion protocol could prevent those who are actually Exposed from mingling with other Susceptible evacuees at the congregate shelter.

\subsubsection*{Dataset formation}
Research has found that people with no symptoms or mild symptoms are the key to understanding the transmission of COVID-19 \cite{gandhi2020asymptomatic, nishiura2020estimation}. Thus, we perturbed the basic reproduction number ($R_0$) (also referred to as the basic reproduction ratio or rate, or the basic reproductive rate), which describes the degree of contagiousness or transmissibility of infectious agents \cite{delamater2019complexity}, to analyze how different vaccine uptake levels and the Diversion protocol delayed the pandemic peak occurrences and reduced the number of infections.

We examined the coupled latency effects of the spread of COVID-19 and the effects of enforcing the Diversion protocol by taking into consideration the following factors: (1) the timing of a mandatory evacuation order announced based on the historical record of debris flows in Kaohsiung City \cite{major}; (2) different brands of vaccines and their efficacy used in Kaohsiung City by September, 2021 \cite{vaccine_kh}; (3) the daily vaccine uptake in the two districts \cite{vaccine_kh}; (4) different population sizes and age distributions of evacuees in each district \cite{populationdata}; and (5) different hypothetical basic reproduction numbers and seeding cases (the initial number of Exposed population).

We also studied various scenarios for evacuating and housing non-symptomatic people (susceptible, exposed, or asymptomatic) versus symptomatic people (mild or severe cases). The Diversion protocol could potentially capture and divert the Exposed (E) evacuees from members of the general public to prevent them from mingling at shelters after a mandatory disaster evacuation was initiated during a COVID-19 outbreak, and thus could decrease COVID-19 mild and severe infections and delay the pandemic peak occurrences. To evaluate the effects of the Diversion protocol in different evacuation conditions, we set up 5 or 10 as the number of seeding cases (the initial number of Exposed people), and $R_0$ was 1.5, 2, 2.5, or 3 in our study.

Since the maximum number of daily vaccinations in the two districts was not publicly available, we assumed that the daily vaccine uptake among the evacuees in each district was in the same ratio as for the total population in Kaohsiung City. Therefore, we used the population size in each district and the historical COVID-19 vaccine doses taken per day in all districts of Kaohsiung City to calculate the proxy number of daily vaccine uptake among evacuees in the two districts . Since vaccine uptake was still quite limited in Taiwan in September 2021, we used the highest level of historical COVID-19 daily vaccine uptake prior to September 1, 2021 in Kaohsiung City \cite{vaccine_kh} to calculate the daily vaccine uptake in each district. After vaccination became widely available in Taiwan, we used the latest COVID-19 daily vaccine uptake as of January 22, 2022 \cite{vaccine_kh} to simulate what if evacuations occurred under sufficient vaccine uptake. By doing so, we were able to study the role of different vaccine uptake levels on the pandemic evolution when evacuees stayed at a congregate shelter for multiple days. As for the vaccine efficacy that we considered in the SEIR simulations, AstraZeneca (AZ) was 70\% effective against infections \cite{chagla2021adults} and Moderna was 95\% effective against infections \cite{mahase2020covid}.

We adapted the SEIR model developed by \cite{hogan2021within} and used the programming language R to investigate the impacts of implementing different vaccine uptake levels and the Diversion disaster evacuation protocol on the evolution of a pandemic. Specifically, we used real-world data, such as the number of evacuees, in our SEIR model simulations. Only those who lived in debris-flow-prone watersheds were listed as evacuees on the registry. Even though the actual number of evacuees in every 5-year group was not available, we assumed the age distribution of evacuees in every 5-year group is the same as the one of the total population in each district \cite{populationdata}. So, we used the total number of evacuees in each district and the ratio of population in every 5-year group to the total population to calculate the proxy number of evacuees in every 5-year-old group in the two districts. We also applied the social contact matrix of Taiwan in the aggregate of every 5-year group, published by \cite{prem2017projecting}.

\subsection*{SEIR model}
\subsubsection*{Model design}
The age-structured SEIR-type mechanistic epidemiological model of COVID-19 dynamics \cite{hogan2021within} used in this study is derived from the basic Susceptible-Exposed-Infectious-Recovered (SEIR) model, which has been widely applied in epidemiology. The conventional SEIR model describes the dynamic evolution of an infectious disease over time and can be expressed by the following coupled ordinary differential equations (ODEs) \cite{heng2020approximately}.

\begin{equation} 
\label{eq:SEIR_S}{\frac{dS}{dt} = -\beta IS} 
\end{equation}

\begin{equation} 
\label{eq:SEIR_E}{\frac{dE}{dt} = \beta IS - \sigma E} 
\end{equation}

\begin{equation} 
\label{eq:SEIR_I}{\frac{dI}{dt} = \sigma E - \gamma I} 
\end{equation}

\begin{equation} 
\label{eq:SEIR_R}{\frac{dR}{dt} = \gamma I} 
\end{equation}
where $t$ is time, $S$ represents Susceptible (uninfected), $E$ signifies Exposed (already infected but not yet infectious so still in the latent period), $I$ means Infectious, and $R$ represents Recovered. Moreover, $\beta = R_0 \gamma$ is the rate of Susceptible becoming Infectious and $R_0$ is the basic reproduction number and $1/\gamma$ is the time Infectious people take to recover. Specifically, $1/\sigma$ describes the latent period, which is the time the Exposed people take to become Infectious. Births and deaths are not considered in the above dynamic process so $S + E + I + R = 1$ \cite{heng2020approximately}.

The main difference between the above basic version of the SEIR model and the SEIR model \cite{hogan2021within} we applied in this study is that we included several vaccination parameters in the dynamic transition from S, E, I, to R. Specifically, the vaccination status transiting from unvaccinated, vaccinated but not protected by the immunity yet, to vaccinated and protected by the immunity, were considered in our SEIR simulations, and these three stages changed over time as sub groups in each compartment (S, E, I, or R) \cite{heng2020approximately}. For more detailed algorithm design and formal mathematical equations of the dynamic process of $S$, $E$, $I$, and $R$ with the vaccination terms, please refer to the Appendix A of \cite{hogan2021within}.

\subsubsection*{Model simulation assumptions}
We made the following assumptions in the SEIR model simulations: (1) Natural immunity from infections lasts one year; this is the default assumption in the original SEIR model \cite{hogan2021within} that we applied and the time window we considered. Thus, Recovered (R) people would not become Susceptible (S) again in the scope of our study, which was less than one year. Moreover, we did not consider breakthrough cases due to the Omicron variants. However, to reflect the influence of new variants, we included scenarios where we increased $R_0$. (2) Each district is a closed environment---no one would enter or leave the district. (3) The vaccine efficacy did not decay within the time window of this study. (4) The SEIR model considers some default vaccination parameters \cite{hogan2021within}, including 2 doses as the default dose schedule (i.e., without the boosters), 7 days as the default mean time to develop the immunity from vaccines after the second dose, 5000 days as the default mean duration of immunity from vaccines, and all ages of the evacuees as the eligible population for the vaccination. (5) We ran the SEIR simulations for the situations where the general public went to a large congregate shelter, like a community center, in each district. When the Diversion protocol was in reality implemented, those who were in self-quarantine were evacuated to non-congregate shelters like hotels, instead of a congregate shelter.

\subsection*{Scenario setting}
Our study constitutes two stages of the SEIR model simulations for the hypothetical debris flow evacuations during a pandemic in the Taoyuan District and the Jiaxian District of Kaohsiung City, Taiwan. 

\subsubsection*{Stage 1}
We ran the SEIR model starting from 5 or 10 seeding cases in each district. Based on the results of the SEIR simulations, we can deduce when the pandemic peaks would occur. $S_t$ and $E_t$ were the number of ``Susceptible (S)'' and the number of ``Exposed (E)'' at time $t$. The evacuation was assumed to occur on the 30th day in Stage 1, which was September 30, 2021. September is in the typhoon season in Taiwan. Thus, evacuations due to rainfall-triggered debris flows occurred several times in Kaohsiung in September according to the debris flow case reports \cite{major}. We used $S_{30}$ and $E_{30}$ to run another set of SEIR model simulations at Stage 2, starting from the moment when a mandatory disaster evacuation occurred during a pandemic.

\subsubsection*{Stage 2}
The purpose of Stage 2 was to simulate four cases where the Diversion protocol can capture and divert different numbers of Exposed who were in self-quarantine from members of the general public who were Susceptible to prevent them from mingling at a congregate shelter after a mandatory disaster evacuation happened during a pandemic.

\begin{enumerate}
\item No Diversion Scenario: all evacuees were evacuated to the same congregate shelter together, and people who were $S_{30}$ and $E_{30}$ mingled at the shelter for multiple days during the disaster recovery process.

\item Diversion Scenario 1: evacuees were evacuated separately, and among $N = 66$ \cite{covid19data} people who were separated from the general public, all of them belonged to ``Exposed (E)'' and none of them belonged to ``Susceptible (S).'' So, ($E_{30}$ - 66) became the seeding cases mingling with $S_{30}$ at the congregate shelter for multiple days. Please note that when $E_{30}$ was less than $N = 66$, which was the number of ``Exposed (E)'' to be deducted from $E_{30}$, $E_{30}$ cannot be negative numbers. Thus, we assumed that the number of infections stayed at 0 throughout the 1-year time window in this study. However, we assumed a nominal pandemic outbreak peak at the end of 365th day in Stage 2 to imply that an outbreak might occur right beyond the 1-year time window in this study.

\item Diversion Scenario 2: evacuees were evacuated separately, and among $N = 66$ people who were separated from members of the general public, 50\% of them belonged to ``Exposed (E)'' and 50\% of them belonged to ``Susceptible (S).'' So, ($E_{30}$ - 33) became the seeding cases mingling with ($S_{30}$ - 33) at the congregate shelter. The same aforementioned assumption applied when $E_{30}$ was less than 33.

\item Diversion Scenario 3: evacuees were evacuated separately, and among $N = 66$ people who were separated from members of the general public, none of them belonged to ``Exposed (E)'' and all of them belonged to ``Susceptible (S).'' So, $E_{30}$ became the seeding cases mingling with ($S_{30}$ - 66) at the congregate shelter. 

\end{enumerate}

\section*{Data availability}
The datasets generated during and/or analyzed during the current study are available from the corresponding author on reasonable request.

\printbibliography
\section*{Acknowledgements}

\textit{Technical Supports:} We thank Dr. Mary McDevitt of the Technical Communication Program in the School of Engineering at Stanford University and Professor Erin Mordecai for rigorously reviewing and revising our manuscript, and providing constructive feedback and valuable instructions in detail.

\textit{Datasets:} We appreciate that Professor Chyan-Deng Jan and Yi-Chao Zeng from National Cheng Kung Univeristy provided us with the number of evacuees on the evacuation registry and the information about the congregate shelters at the Taoyuan District and the Jiaxian District of Kaohsiung City, Taiwan.

\textit{Funding:} This research was supported by Stanford Woods Institute for the Environment, Department of Civil and Environmental Engineering at Stanford University, Stanford RISE (Respond. Innovate. Scale. Empower.) COVID-19 Crisis Response Research Grant and Fellowship, and National Center for High-performance Computing (NCHC), Taiwan.

\section*{Author contributions}
All authors designed, developed, and refined the methodologies of analyses; Y.L.T. and D.Y.S. performed the analyses; all authors interpreted results. Y.L.T. wrote the first draft of the manuscript. All authors revised the manuscript and reviewed the final version of the manuscript.

\section*{Ethics declarations}
\subsection*{Competing interests}
The authors declare no competing interests.

\section*{Disclaimer}
This disclaimer informs readers that this study should only serve as their own references rather than official guidelines for any jurisdictions. The scenarios of simulations performed in this study are only examples for the purpose of academic research. Any actions, including but not limited to decisions, policies, and studies, taken based on any part of this study is the sole liability of readers, not authors in this study.

\end{document}